\documentstyle[prl,aps,psfig,twocolumn]{revtex} 

\newcommand{\beq}{\begin{equation}}
\newcommand{\beqar}{\begin{eqnarray}}
\newcommand{\eeqar}{\end{eqnarray}}
\newcommand{\beqars}{\begin{eqnarray*}}
\newcommand{\eeqars}{\end{eqnarray*}}
\newcommand{\eeq}{\end{equation}}
\begin{document}
\draft
\twocolumn[\hsize\textwidth\columnwidth\hsize\csname
@twocolumnfalse\endcsname
\title{Directed Polymers on a Factorized Disorder Landscape}
\author{Paolo De Los Rios and Yi-Cheng Zhang}
\address{Institut de Physique Th\'eorique, 
Universit\'e de Fribourg, CH-1700, Fribourg, Switzerland.}
\date{\today}
\maketitle

\begin{abstract}
We study  the Directed Polymer model subject to a particular
form of disorder, $\eta(x,t)=\eta_X(x) \eta_T(t)$, recently proposed
in biological applications.
We find that two new universality classes arise, depending on 
the the lattice geometry.   
Using an intermediate model linking the two different
orientations continuously, we find that there is a phase transition 
separating two distinct scaling phases. For both phases we
get a reasonable understanding of the nature and values of the 
exponents, corroborated with numerical results.
\end{abstract}
]
\narrowtext

In their study of DNA-sequence alignments, Hwa and L\"assig
consider the mapping of the problem of sequence recognition and matching
to a more standard (in statistical
mechanics) model of directed polymers (DP) subject to a special 
disorder~\cite{HL96}. They establish, both
analytically and numerically, the equivalence of the two problems.
Indeed, their model is extremely similar to the DP problem
~\cite{HHZ95}
but with a special disorder structure. Whereas in the usual
DP problem every site $(x,t)$ on the lattice has an energy $\eta(x,t)$
independent from the energy of any other site, Hwa and L\"assig
define $\eta(x,t)$ as
\begin{equation}
\eta(x,t) = \eta_X(x) \eta_T(t)\;\;,
\label{factorized disorder}
\end{equation}
where $\{ \eta_X(x) \}$ and $\{ \eta_T(t) \}$ are independent unbiased
({\it i.e.} with zero average) random variables. 
For the  biological motivations to consider this kind of disorder 
we refer the reader to Hwa and L\"assig's work\cite{HL96}. 
Naively one may expect that this disorder distribution should give
no difference with respect to the usual uncorrelated disorder. Indeed
correlations may appear non-relevant, since
\begin{equation}
\overline{\eta(x_1,t_1) \eta(x_2,t_2)} = \overline{\eta_X(x_1) \eta_X(x_2)}
\;\; \overline{\eta_T(t_1) \eta_T(t_2)}
\label{first correlator}
\end{equation}
is zero if $ x_1 \ne x_2$ and $t_1 \ne t_2$, as for the usual uncorrelated
disorder. However, the count of the total number
of disorder degrees of freedom suggests that the model should be very different.
On a $2$-dimensional square lattice of 
linear size $L$ ($L^2$ sites) there are 
only $2L$ degrees of freedom, 
versus $L^2$ degrees of freedom for the usual site disorder. 
In other contexts this difference leads to relevant and important
consequences. Indeed, we recall the Mattis model
of spin-glasses \cite{Mattis76}, where spin coupling disorder is replaced by 
a product of disorders, and frustration effects disappear.

In this Letter we systematically study the effects of such a factorized disorder
on the properties of directed polymers, both with numerical simulations
and with theoretical arguments. We find that two new universality
classes emerge, depending on the orientation of the lattice. We
then show that it is possible to obtain both universality
classes on a same system by varying a parameter, a signature of 
a phase transition. Finally, we give explanations
for the difference between our results and Hwa and L\"assig's 
ones\cite{HL96}.

Eq.(\ref{factorized disorder}) defines the model on a rectangular lattice
with disorder defined along the $(1,0)$ and $(0,1)$ directions. 
We can also define the model on the diagonal $(1,1)$ and $(-1,1)$ directions, 
i.e. using the so-called diamond lattice.
In the latter case Eq.(\ref{factorized disorder}) becomes
\begin{equation}
\eta(x,t)=\eta(t+x) \eta(t-x)
\label{diamond disorder}
\end{equation}
as was originally studied in \cite{HL96}.

First we study numerically the model on a diamond lattice, 
eq.(\ref{diamond disorder}). 
As for the usual study of the ground state of directed polymers
({\it i.e.} temperature $T=0$), we are 
interested in the scaling properties
of the polymers as a function of their length. 
In particular we consider the
transverse wandering fluctuations of the polymer, 
$<|\delta x|> \sim t^\nu$, where angular
brackets indicate the average over different realizations 
of disorder (as usual in
quenched disorder problems). Also, the ground state 
energy fluctuations are of interest,
$<|\delta E|> \sim t^\omega$. 

We always use disorder taken from uniform distributions over compact supports,
in general in the interval $[-1/2,1/2]$. We discuss later the effects
of biased distributions.

A summary of the exponents obtained is given in Table \ref{tab1}.

\begin{figure}
\centerline{\psfig{file=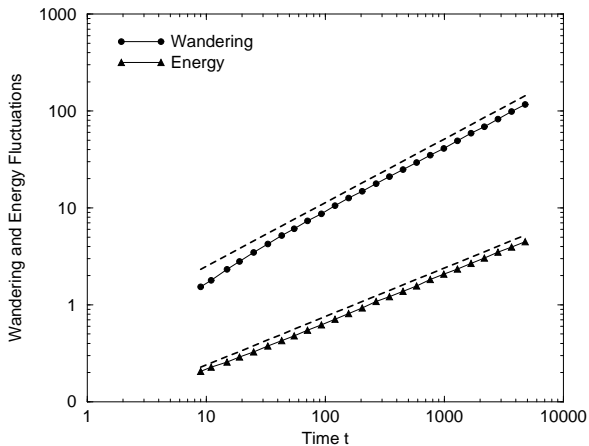,width=8.5cm,angle=0}}
\caption{Wandering and Energy Fluctuations for polymers on the diamond lattice;
data are obtained after averaging over $10000$ disorder realizations.}
\label{Fig: fig1}
\end{figure}

For the $d=1+1$ diamond lattice we observe indeed the scaling of $\nu=2/3$,
consistent
with the standard disorder result~\cite{HHZ95}, see Fig.\ref{Fig: fig1}. Yet,
the ground state energy fluctuations scale with an exponent $\omega=1/2$
different from the standard $\omega = 1/3$. 
In order to appreciate the full mechanism 
of the model, and to understand where the difference comes from,
we study the time evolution of the energy profiles on diamond lattice
strips of various width $L$. It corresponds to the growth of a 
KPZ-like~\cite{HHZ95,KPZ86} surface in discrete space.

The beahvior of the energy profile fluctuations 
$\Delta E(L,t) = <\overline{(E(x,t)-\overline{E(x,t)})^2}>^{1/2}$
(over-bars represent averages over one single 
disorder realization)
is governed by three exponents. Initially $\Delta E(L,t)$ grows with time
as $\Delta E(L,t) \sim t^\beta$. Then it saturates to a constant value,
dependent on the size $L$ of the system as  $\Delta E(L,t) \sim L^{\chi}$.
The characteristic time
$\tau$
separating the two regimes scales with the size $L$ of the system as
$\tau \sim L^z$, and $z$ is called dynamical exponent.
Scaling 
consistency imposes $z = \chi/\beta$.
In standard DP-KPZ theory, the surface exponents and the polymer
ones are related by $z=1/\nu$ and $\beta = \omega$.
Moreover, in standard KPZ theory, the law $z+\chi=2$ holds, due
to galilean invariance.

The scaling analysis of our simulations gives
$\chi=1/2$, as in standard KPZ theory, and $\beta=1/2$. 
From scaling consistency we have $z= \chi/\beta =1$, which is in agreement 
with numerical simulations, but that does not satisfy neither the relation
$z=1/\nu = 3/2$, nor $z+\chi=2$.
The reason behind the above scaling behavior, namely $z=1$ instead of
$z = 1/\nu = 3/2$, is due to the particular structure of the disorder
landscape. Indeed, the diagonal $(\pm 1,1)$ 
factorization of disorder imposes a 
space-time linear proportionality relation, 
corresponding to a dynamical exponent
$z=1$. We therefore find that the usual relation $z = 1/\nu$
breaks down. Let us now try to explain the $1/2$ value of the $\chi$ and 
$\omega=\beta$ exponents. Indeed the $\omega=1/2$ could come from 
an uncorrelated
random walk, whereas $\chi=1/2$ is consistent
with the standard result. In order to solve this ambiguity we have analyzed
the energy profile evolution for a $2+1$ dimensional diamond lattice.
Again we find $z=1$, strengthening the time-space proportionality
interpretation, and $\beta = \chi = 0.40 \pm 0.01$, nicely consistent
with the standard result~\cite{HHZ95,AF90}. We conclude therefore 
that the 
asymptotic behavior
is KPZ dominated. At variance with standard KPZ, the approach to the
stationary state changes, governed by a {\it ballistic} dynamical
exponent $z=1$.

\begin{figure}
\centerline{\psfig{file=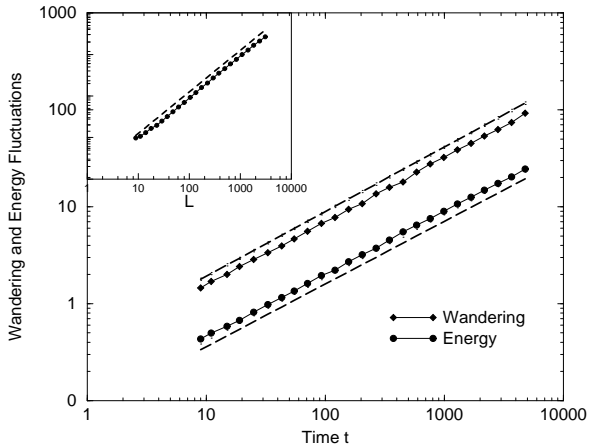,width=8.5cm,angle=0}}
\caption{Wandering and Energy Fluctuations for polymers on the diamond lattice;
data are obtained after averaging over $10000$ disorder realizations. Roughness
of the energy profile on strips of different width $L$, again averages
are over $10000$ samples.}
\label{Fig: fig2}
\end{figure}

Let us now consider the square lattice with disorder defined from
Eq. (\ref{factorized disorder}). 
The directed polymer scaling results are shown in Fig.\ref{Fig: fig2}.
The exponents are consistent with $\nu = 2/3$, as in the diamond case,
and $\omega = 2/3$.
From energy profile evolution, we find the exponents 
$\beta=2/3$ and $z=3/2$ (consistent respectively with $\beta = \omega$
and $z=1/\nu$). The roughness exponent is then 
$\chi = z \beta = 1$, in agreement with numerical results.
The absence of any space-time relation imposed by the disorder structure
leaves therefore the usual $z=1/\nu$ relation unchanged. 

\begin{figure}
\centerline{\psfig{file=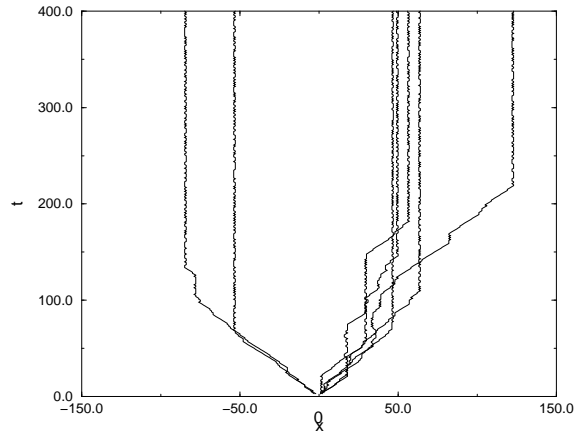,width=8.5cm,angle=270}}
\caption{Ground state polymers on a square lattice. 
Every polymer corresponds to a different disorder realization.}
\label{Fig: fig3}
\end{figure}

We explain the value $2/3$ of the $\nu$ and $\omega$ exponents using 
an argument generalized from
columnar disorder theory~\cite{HHZ95,Z86}. Indeed, as emerges from 
Fig.\ref{Fig: fig3},
after an initial wandering, polymers tend to localize in a very narrow
region of the lattice, in general consisting
of a pair of columns. The corresponding two $\eta_X(x)$ values have 
opposite signs and their absolute values are extremely close to $1/2$.
Polymers are then attracted to such pairs, 
because they can jump from one column to the other depending
on the sign of $\eta_T(t)$, always choosing the negative product 
$\eta_X(x) \eta_T(t)$ in their quest for low energies.
We can therefore think that polymers search the best pairs available.
The relevant feature of the system is therefore the effective pair 
energy probability distribution $Q(\epsilon)$. From its knowledge, 
using variational
arguments, it is possible to obtain an analytical estimate
of the exponents.
Indeed, since the polymer always
chooses the negative product, as long as we are interested in an {\it effective}
pair energy distribution  
we can disregard the $\eta_T(t)$ energies, just remembering that the
chosen $\eta_X(x)$ energies are extremal ({\it i.e.}, as close as possible to
$\pm 1/2$). Moreover a non-restrictive extra condition 
is that the two $\eta_X(x)$ energies are equal in absolute value. This is  
numerically well approximated, since the two energies are both extremely close,
in absolute value, to $1/2$.

Let us define $\epsilon$ such that the energies of the two $\eta_X(x)$
of the pair are $\pm (1/2-\epsilon)$. Then $\epsilon \in [0,1]$.
The probability that the pair energies are closer than $\epsilon$ to $\pm 1/2$
is
\begin{equation} 
\int_0^\epsilon Q(y) dy = \left[ \int_{-\frac{1}{2}}^{-\frac{1}{2}+\epsilon }
dy \right] \; \left[ \int_{\frac{1}{2}-\epsilon}^{\frac{1}{2}}
dy \right]  = \epsilon^2
\label{pair integr prob}
\end{equation}
leading to a probability distribution for $\epsilon$ that is 
$Q(\epsilon) = 2 \epsilon$.
$Q(\epsilon)$ is then the relevant pair {\it effective} distribution.

It is then possible to apply a variational argument to give
a prediction on the exponent $\nu$. Indeed, the polymer wanders initially
in a region of characteristic width $R$, in search of the optimal (pair)
energy. Once it has found the optimal pair, it localizes there up
to a time $t$. The global energy can therefore be evaluated as
\begin{equation}
E = \gamma R + (t-R) \epsilon_{min}
\label{variational estimate}
\end{equation}
The optimal energy available in a region of width $R$ can be obtained from
the statistics of the extrema. Indeed, the probability $Q_m(\epsilon)$ 
of the smaller energy among $R$ energies distributed according
to $Q(\epsilon) = (\mu +1) \epsilon^\mu$ in the interval $[0,1]$ is
\begin{eqnarray}
Q_m(\epsilon) = &R& Q(\epsilon) \left[1 - \int_0^\epsilon Q(y) 
dy \right]^{R -1} = \\ 
&R& (\mu+1) \epsilon^\mu \left( 1 -\epsilon^{\mu+1} \right)^{R-1}
\label{extremum probability}
\end{eqnarray}
Then the characteristic extremal energy $\epsilon_{min}$ is 
\begin{equation}
\epsilon_{min} = \int_0^1 \epsilon Q_m(\epsilon) d\epsilon = R\; B\left(R-1,
\frac{\mu+2}{\mu+1}\right)\;\;,
\label{characteristic extremal}
\end{equation}
where $B(x,y)$ is the Euler Beta function.
After some manipulation, the scaling behavior of $\epsilon_{min}$ is found to
be $\epsilon_{min} \sim R^{-1/(\mu+1)}$ for large $R$. 
Inserting this result in
(\ref{variational estimate}) and optimizing with respect to $R$, we find the
scaling behavior $R \sim t^{(\mu+1)/(\mu+2)}$ and therefore 
$\nu=(\mu+1)/(\mu+2)$; in our case $\mu=1$ and 
therefore $\nu = 2/3$, as from numerical simulations. Also, the energy
$E$ scales as $R$ and therefore $\omega=\nu$, again in agreement
with numerical simulations. 

We can also change the probability distribution of the energies
along the $x$ and $t$ directions. As long as both energies can take on
positive and negative values, and the polymer does not loose 
any energy wandering through different columns, 
the universality class does not change.
It does not even change if the energies $\eta_X(x)$ take only positive values:
the polymers continue to localize on pairs of energies, to maximize the energy
gain when the $\eta_T(t)$ energy is negative and to minimize the energy loss
when $\eta_T(t)$ is positive. The only different universality class
comes when the $\eta_T(t)$ energies assume only positive values: in that case
there is no more any convenience for the polymers to bind to a pair, 
and localization takes place over a single optimal $\eta_X(x)$ energy.
Being the energy distribution flat ($\mu=0$), the exponents
are $\nu = \omega = 1/2$, in agreement with numerical results.   
If instead there is an energy cost to change column, then a transition 
as a function of the bias in the distributions takes place between the two
scaling regimes $\nu = 1/2$ and $\nu=2/3$. The full phase diagram 
of the latter situation will be published elsewhere. 

This polymer problem defined by (\ref{factorized disorder}) is indeed 
sensitive to the chosen disorder distribution. For instance, using 
Gaussians distributions, the variational procedure predicts a 
linear relation between $R$ and $t$ 
(apart from logarithmic corrections~\cite{Z86}), confirmed
by our numerical simulations.

Next, we move on to show that the two different scaling behaviors (namely,
$z=1$ and $z=3/2$) are
compatible on the same disorder configuration. Let us start from the square
lattice: a polymer  
has three possibilities to go one step forward, namely, to the left, in front,
to the right.
In our simulations,
we chose total isotropy, that is, neither of the three 
choices imply any energy loss, and the $z=3/2$ result
is recovered. We let then one of the directions,
say the leftward one, be energetically unfavorable, with an associated 
energy cost $a$. Consequently the polymer end-points show
a drift in time toward the right direction. Indeed, 
we find that, for very large values of $a$, the polymer
average final position moves to the right linearly with time. 
The fluctuations of the
end-points around their average position scale with time with exponent
$\nu = 2/3$, and the energy fluctuations scale with exponent $\omega = 1/2$,
as for the diamond lattice case.
Indeed, looking at the scaling properties of the energy profile, 
we recover the full set of exponents typical of polymers with factorized
disorder on the diamond lattice, Eq.(\ref{diamond disorder}).
Strong anisotropy induces therefore a linear space-time relation
(as emerges from the linear drift of the polymer end-points)
that forces a $z=1$ dynamical exponent.
For small
values of $a$ (down to the isotropic case $a=0$), 
instead, we find the other full set of exponents, typical of
disorder (\ref{factorized disorder}). 

\begin{figure}
\centerline{\psfig{file=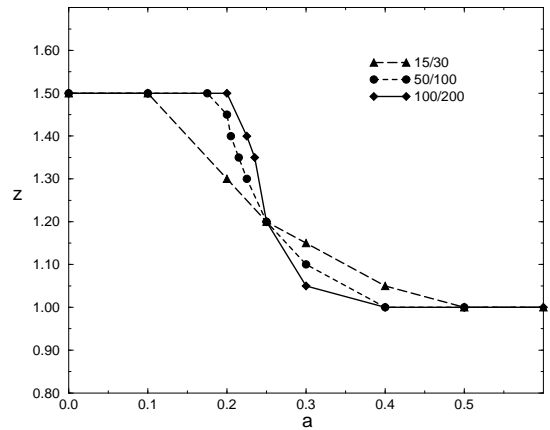,width=8.5cm,angle=270}}
\caption{Exponent $z$ vs. anisotropy energy $a$, as obtained from collapse plots
of roughnesses for different strip widths.}
\label{Fig: fig4}
\end{figure}

As Fig.\ref{Fig: fig4}
shows, by varying $a$ we find the signature of a phase transition between the
two regimes $z=3/2$ and $z=1$. The three point sets are obtained from collapse
plots of the scaling features of energy surfaces for different system sizes. 
Precisely, the values plotted come from the collapse of data of couples of
strips. The values of the width pairs are given in the legend.
The data are steeper with increasing strip width 
in the neighbor of $a=0.25$. We expect $a=0.25$ to be
precisely the transition value of the anisotropic energy. Indeed, 
localization takes place when, on the average, it is convenient for
polymers to localize even if it has to pay a price $a$: a localized polymer
chooses the column of the two where it
finds negative disorder energies. Since the sign of $\eta_X(x)$ is
fixed, it finds a sequence of $n$ negative values on the same column
if $\eta_T(t)$ keeps the same sign $n$ times, and this happens with
probability $(1/2)^n$. The average length $\bar{n}$ of a sequence with
the same sign is therefore 
\begin{equation}
\bar{n} = \sum_{n=1}^{\infty} n \left(\frac{1}{2}\right)^n = 2 \;\;.
\label{average}
\end{equation}
The average energy $|\overline{\eta_T(t)}| = 1/4$ (the sign depends on the sign of
$\eta_X(x)$). The absolute value of the energy $\eta_X(x)$ 
is extremely close to $1/2$ 
(again irrespective
of the sign, which depends on the sign of $\eta_T(t)$ so that 
their product is
negative).
The average energy that a localized polymer gains against $a$ is therefore
\begin{equation}
\bar{\epsilon} = -\frac{1}{4} \frac{1}{2} \bar{n} = - \frac{1}{4}\;\;.
\label{average gain}
\end{equation}
Therefore if $a<1/4$ localization takes place.
Such a transition was also alluded to in Ref.\cite{HC96+}.

Summarizing, we have found that a factorized disorder distribution
such as (\ref{factorized disorder}) or (\ref{diamond disorder}) leads
to new universality classes, depending on the orientation of the lattice.
We have been able to explain the origin of the characteristic exponents
of these classes and to connect them to each other via an
anisotropy driven phase transition. 

A challenging issue is to connect these results to Hwa {\it et al.}
~\cite{HL96,HC96+}, where the usual disorder universality class 
was obtained~\cite{HHZ95}. Indeed, in their work
they defined the disorder and the lattice in a slightly different fashion
(see Refs.\cite{HL96,HC96+} for details).
As we showed above, the polymer problem with factorized disorder
is very sensitive to the details of the disorder and of the lattice.
We believe that the reason underlying
this difference resides in the presence, in their work, of some 
directions along which polymers do not feel any disorder (the {\it gaps}
in the language of Refs.\cite{HL96,HC96+}). 
Indeed, depending on the energy cost of these directions, 
we found indications of a phase transition between the standard
disorder universality class and the diamond lattice one 
(see Table \ref{tab1}). Further work is needed in this direction,
to explore the full phase diagram of the model.

\begin{table}[ht]
\begin{center}
\begin{tabular}{c|ccc}
$disorder$       & Diamond    & Square   & Standard  \\
\hline
$\nu$            & 2/3 & 2/3 & 2/3  \\
$\omega=\beta$           & 1/2 & 2/3 & 1/3 \\
$z$         & 1 & 3/2 & 3/2   \\
$\chi$         & 1/2 & 1 & 1/2  \\
\end{tabular}
\caption{Characteristic exponents for directed polymers and energy
surfaces for the diamond lattice (disorder defined from Eq.(3)),
square lattice (Eq.(1)), and for the standard disorder [3]. }
\label{tab1}
\end{center}
\end{table}
\end{document}